\documentclass[conference]{IEEEtran}

\usepackage[top=1in,bottom=0.75in,right=0.75in,left=0.75in]{geometry}

\usepackage[algo2e]{algorithm2e}

\usepackage[noend]{algpseudocode}
\usepackage{pifont}
\usepackage[utf8]{inputenc}
\usepackage[noadjust]{cite}
\usepackage{amsmath}
\usepackage{times}
\usepackage{textcomp}
\usepackage{amsfonts}
\usepackage{amssymb}
\usepackage{amsthm}
\usepackage{graphicx}
\usepackage{float}
\usepackage{layout}
\usepackage{array}
\usepackage{comment}
\usepackage{array}
\usepackage{color}
\usepackage[usenames,dvipsnames]{xcolor}
\usepackage{soul}
\usepackage{footmisc}
\theoremstyle{plain}
\usepackage{epstopdf}
\usepackage{amssymb}
\usepackage{bbm}
\usepackage{mathtools}
\usepackage{cuted}
\usepackage{lipsum, color}
\usepackage{tikz}
\usepackage{tkz-tab}
\usetikzlibrary{automata,arrows,positioning,calc}
\usepackage{caption}
\usepackage{subcaption} 

\usepackage{dsfont}

\newtheorem{claim}{Claim}

\newtheorem{example}{Example}

\renewcommand{\vec}[1]{\mathbf{#1}}
\newcommand{\mI}{\mathbb{I}}

\DeclareMathOperator\erf{erf}
\makeatletter
\newcommand*{\rom}[1]{\expandafter\@slowromancap\romannumeral #1@}
\makeatother
\usepackage{graphicx}
\graphicspath{ {images/} }
\usepackage{amssymb}

\usepackage{amsmath}
\usepackage{algorithm}
\usepackage[noend]{algpseudocode}
\makeatletter
\def\BState{\State\hskip-\ALG@thistlm}
\makeatother

\IEEEoverridecommandlockouts
\begin{document}        
%
\title{Towards Mobility-Aware Proactive Caching \\ for Vehicular Ad hoc Networks}

\author{\IEEEauthorblockN{Yousef AlNagar${}^{\star}$, Sameh Hosny${}^{\star}$, Amr A. El-Sherif${}^{\star}$${}^{\mp}$}
\IEEEauthorblockA{${}^{\star}$ Wireless Intelligent Networks Center (WINC), Nile University, Giza, Egypt.
\\${}^{\mp}$ Department of Electrical Engineering, Alexandria University, Alexandria 21544, Egypt.
\\
Emails: Y.Alnagar@nu.edu.eg, SHosny@nu.edu.eg, aelsherif@nu.edu.eg}
}

\maketitle
\begin{abstract}
Harnessing information about the user mobility pattern and daily demand can enhance the network capability to improve the quality of experience (QoE) at Vehicular Ad-Hoc Networks (VANETs). Proactive caching, as one of the key features offered by 5G networks, has lately received much interest. 
However, more research is still needed to convey large-sized
multimedia content including video, audio and pictures to the high speed moving vehicles. In this paper, we study the gains achieved by proactive caching in Roadside Units (RSUs) where we take into consideration the effect of the vehicle velocity on the optimal caching decision. Information about the user demand and mobility is harnessed to cache some files in RSUs, which will communicate with vehicles traversing along the visited roads before the actual demand. Our main objective is to minimize the total network latency. Towards this objective, we formulate two optimization problems for non-cooperative and cooperative caching schemes to find the optimal caching policy to decide which files to be cached by the RSUs. Due to the complexity of these problems, we propose a sub-optimal caching policy for each scheme. We compare the performance of the optimal caching policy to that of the sub-optimal caching policy.
Numerical results show that proactive caching has a significant performance gain when compared to the baseline reactive scenario. Moreover, results reveal that the cooperative caching scheme is more efficient than the non-cooperative scheme.  
\end{abstract}
\IEEEpeerreviewmaketitle
\section{Introduction}

Global mobile data traffic reached 96 exabytes per month at the end of 2016 and is expected to reach 278 exabytes per month by 2021. The proliferation of online social networks leads to an explosion of video traffic which is expected to reach 80\% of the total internet traffic \cite{paper_2}. Users demand increases dramatically and network traffic grows exponentially. Therefore, communication latency becomes a vital issue to consider. A promising approach to meet these challenges at cellular networks is caching at the network edge to mitigate overload on the remote backhaul servers \cite{paper_3}. In caching, the most frequent files are accessed from some intermediate nodes to reduce load on the servers. As user demand is not processed by the server, there is less time delay in retrieving the service. Today there is significant consensus that caching will play a fundamental role in the future communication systems \cite{paper_20}. In particular, \emph{proactive caching} is considered one of the effective solutions in 5G networks. It relies on caching desired data before actual demand which can enhance traffic offloading, energy and spectrum efficiency \cite{paper_6}. However, preserving cache information, in order to provide the moving users with the requested data items, is a major challenge. Great efforts are pursued to look over mobility consequences on caching systems \cite{paper_4}, \cite{paper_5}.

Vehicular Ad-Hoc Networks, special wireless Ad-Hoc networks in which communication nodes are moving vehicles, are considered one of the most prominent areas for research and industry applications. VANETs depend mainly on Vehicle-to-Vehicle (V2V) and Vehicle-to-Infrastructure (V2I) communications to support abundant applications such as intelligent transportation, emergency services, self driving, information sharing and entertainment \cite{paper_1}. Caching popular files in RSUs with large storage capacity is studied in \cite{paper_7}. Authors try to tackle latency problem where the main objective is to minimize the mean time for On-Board Units (OBU) to download a file. They propose three algorithms to assign files at RSUs, an optimal algorithm that achieves the best performance and depends on exhaustive search, a sub-optimal algorithm, and a low-complexity greedy algorithm. On the other hand, a split content caching scheme is addressed in \cite{paper_8} with an area controller located in the network backhaul which manages caching placement in the RSU layer. The authors present a simple analytic model to predict the probability for the content to be requested at particular edge nodes.

Cooperative content caching between moving vehicles is introduced in \cite{paper_9}. The outage probability is considered as a performance metric for the proposed caching model. The authors also discuss the outage analysis under random mobility scenario, and vehicular scenario where vehicles are assumed to move in straight freeways. Authors in \cite{paper_10} present a caching placement policy to maximize the caching gain based on the cloud-based VANET at the vehicular layer and RSU. Vehicles are assumed to travel in platoon and RSUs are deployed uniformly along the road. A new caching allocation policy is suggested to jointly consider the vehicle and RSU caches. An optimization problem is formulated to minimize the latency for vehicles to receive their desired data items. Although previous work studied convenient caching schemes at both vehicles and RSUs, they didn't capture the effect of vehicles' velocity and how to exploit users' mobility patterns to
enhance the system performance. There is a strong evidence in the literature that many people follow certain daily routines and hence their behavior (mobility pattern and demand) is highly predictable over the time as a prior information \cite{paper_12,paper_13,paper_14,paper_19}. In this work, we principally focus on deploying a cooperative proactive caching scheme between RSUs to minimize the communication latency and enhance quality of service (QoS) in VANETs. We show that exploiting the information about user mobility and demand statistics helps us to address the latency problem and reinforces the network capacity. The main contributions are:
\begin{enumerate}
	\item We propose a non-cooperative caching scheme where each RSU finds its optimal caching decision independently. The optimization problem is formulated in order to minimize the total expected delay of the network. A reactive network is considered as a baseline model to evaluate the performance of the proposed model.
	\item Due to the complexity of the optimization problem, we introduce a sub-optimal caching policy which achieves a high caching gain compared to the baseline reactive scheme.
	\item We extend our work by considering a cooperative caching scheme where RSUs collaborate to get their optimal caching decisions after considering the vehicles updated demand and location information. We propose a greedy algorithm which outperforms the non-cooperative caching scheme. We show that cooperation between RSUs enhances the network performance and minimizes the total expected delay.
\end{enumerate}

The rest of the paper is organized as follows. In Section \ref{Sec:System_Model}, we introduce the system model and state the underlying assumptions. The problem is formulated in Section \ref{Sec:Problem_Formulation}. We discuss the non-cooperative caching scheme in Section \ref{Sec:NonCooperative}. In Section \ref{Sec:Cooperative}, we propose the cooperative caching scheme and introduce a sub-optimal caching policy. Numerical results and discussions are provided in Section \ref{Sec:Results}. The paper is concluded in Section \ref{Sec:Conclusion}.
\section{System model}\label{Sec:System_Model}
\begin{figure}
        \centering
        \includegraphics[width=0.45\textwidth]{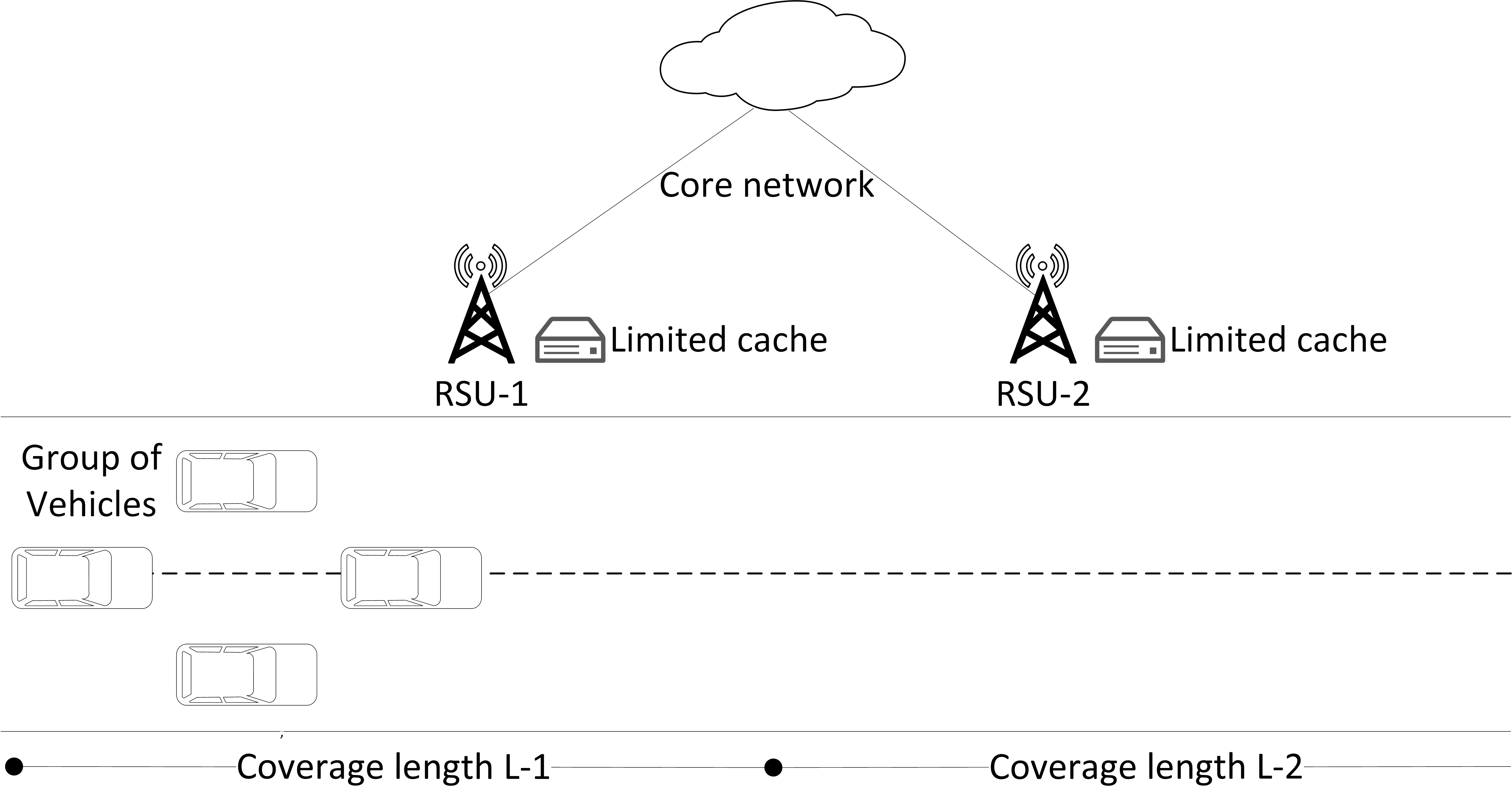}
        \caption{System Model.}
        \label{fig:1}
    \end{figure}

We consider a set of $S$ Roadside Units, $\mathcal{S}=\{1,2,\cdots,S\}$, that are located equidistantly with distance $L_s$ over a certain road, as shown in Fig \ref{fig:1}. Each RSU $s\in \mathcal{S}$ is equipped with a limited cache of size $Z_s$ which is used to cache a number of data items. Along the road, there is a set of $V$ moving vehicles,  $\mathcal{V}=\{1,2,\cdots,V\}$. Vehicles are interested in a set of $M$ uncorrelated data items,  $\mathcal{M}=\{1,2,\cdots,M\}$. For simplicity, we assume that all data items have the same size of $C$ bytes. Each RSU $s$ serves connected vehicles, which are moving within its coverage area, with data rate $\alpha_s$ bytes/sec. In the free-flow state when the traffic density is low, we assume that vehicles velocities are independent and identically distributed since drivers can choose their appropriate speeds, which are generated randomly by truncated Gaussian distribution and each vehicle keeps its assigned speed $u_v$ while it moves along the highway \cite{paper_22},\cite{paper_23}. It is widely acceptable that velocity distribution of vehicles in freeways follows Gaussian distribution \cite{paper_21}. Without loss of generality, we assume that any vehicle is moving with a velocity $u$ that is limited by minimum and maximum speeds, i.e. $u_{min} \leq u \leq u_{max}$. Therefore, we have truncated normal distribution with mean $\mu$ and variance $\sigma^2$ where
    \begin{equation}\label{Eq:truncated_exponential_distribution}
            f(u) = \begin{cases}
                \frac{2\exp \left(\frac{-(x-\mu)^2}{2\sigma^2} \right)}
                {\sqrt{2\pi \sigma^2}\left(\erf \left(\frac{u_{max}-\mu}{\sigma \sqrt{2}}\right)-\erf \left(\frac{u_{min}-\mu}{\sigma \sqrt{2}} \right) \right) },&\\ \hspace{33mm}u_{min} \leq u \leq u_{max},\\
                0,&\hspace{-31mm} \text{otherwise.}
            \end{cases}
    \end{equation}
Let $\theta_v^s$ be the probability that vehicle $v$ is located at the coverage area of RSU $s$. Due to their mobility, vehicles will be connected to any RSU for a certain amount of time and then handover to the next RSU. We define the contact time vector $\vec{h}^s:=\left(h_{1}^s,h_{2}^s,\cdots,h_{v}^s\right)$ where $h_{v}^s$ is the contact time between vehicle $v$ and RSU $s$ in seconds.
We assume that the RSUs can track, learn and predict the vehicle behavior, hence,  construct a \emph{demand profile} for each vehicle $v$ denoted by $\vec{p}_{v}:=\left(p_{v}^{1},p_{v}^{2},\cdots,p_{v}^{M}\right)$ where $p_{v}^{m}$ is the probability that vehicle $v$ requests data item $m$. The demand of vehicle $v$ is captured by a random variable expressed by
	\begin{equation*}\label{Eq:Demand_Indicator}
	\mI_{v}^{m}=\begin{cases}1, & \text{with probability } p_{v}^{m},\\
	                     0, & \text{with probability } 1-p_{v}^{m},
	\end{cases}
	\end{equation*}
where $\mI_{v}^{m}=1$ means that data item $m$ is requested by vehicle $v$ while $\mI_{v}^{m}=0$ means that it is not requested. We assume that $\mI_{v}^{m_1}$ is independent of  $\mI_{v}^{m_2}, \forall m_1,m_2 \in \mathcal{M}$. Moreover, for any $v_1 \neq v_2$, $\mI_{v_1}^{m}$ is independent of $\mI_{v_2}^{m}, \forall m \in \mathcal{M}, \forall v_1, v_2 \in \mathcal{V}$. Let $x_s^m $ be the caching decision of data item $m$ at RSU $s$ where
   \begin{equation}\label{Eq:Caching_Const_1}
        x_s^m \in \{0,1\},\forall m \in \mathcal{M}, \forall s \in \mathcal{S}.
    \end{equation}
In particular, $x_s^m=0$ means that the RSU does not cache data item $m$ while $x_s^m=1$ means the data item $m$ is cached at RSU $s$. It is worth to notice that the size of the cached items should be less than or equal to the total RSU storage $Z_s$.
Therefore, we have
    \begin{equation}\label{Eq:Caching_Const_2}
        C \sum_{m=1}^{M} x_s^m \leq Z_s, \forall s \in \mathcal{S}.
    \end{equation}
When a vehicle requests a data item, the serving RSU transmits the requested data item to the vehicle, in $\frac{C}{\alpha_s} $ seconds, if the data item is available in its local cache. If not found locally, the RSU requests this data item from the network backhaul. In this case, every data item $m$ requires time $(\tau+ \frac{C}{\alpha_s} )$ seconds to be served, where $\tau$ is the time to fetch the requested data item from backhaul to the RSU. We aim to find an optimal caching policy $\{ x_s^{m*} \}$ that specifies the data items to be be cached at the RSU to minimize the total latency.


\section{problem formulation}\label{Sec:Problem_Formulation}

RSUs take the vehicles mobility and demand statistics into consideration to find an optimal caching policy. To study the potential gain achieved by the proactive caching approach, we compare the proposed model with a baseline reactive scheme. Let $h_v$ represents the contact time between the vehicle $v$ and the serving RSU $s$ where
    \begin{equation}\label{Eq:Const_General_Reactive}
        h_v^{s} = \frac{L_s}{u_v}. 
    \end{equation}

\subsection{Reactive Network}

We consider a \emph{reactive network} as a baseline scenario where the requests are served directly from the network backhaul without caching at the RSUs. Let $M_v^s$ be the maximum number of data items that can be received by vehicle $v$ from RSU $s$, where 
    \begin{equation}
        M_v^s := \min \left( \left\lfloor{ \frac{h_v^s}{\frac{C}{\alpha_s}+\tau} }\right\rfloor,M \right).
    \end{equation}
Each data item will be delivered from the network backhaul within ($\frac{C}{\alpha_s} + \tau$) seconds. Hence, each vehicle can receive up to $\left\lfloor{ \frac{h_v^s}{\frac{C}{\alpha_s}+\tau} }\right\rfloor$ data item but can not exceed the library size $M$. In particualr, each vehicle can request $k$ data items where $0 \leq k \leq M_v^s$. Therefore, the expected total delay incurred by all vehicles for RSU $s$ will be given by 
    \begin{equation}\label{Eq:reactive_time}
        W_{s}^\mathcal{R} = \sum_{v=1}^{V} \left( \theta_v^{s} \sum_{k=0}^{M_v^s} k \sum_{i=1}^{G_k} q_{i,k} \left( \frac{C}{\alpha_s} + \tau \right)\right),
    \end{equation}
where $\mathcal{R}$ represents the reactive network and $G_k=\binom{M}{k}$ is the number of all possible combinations to request $k$ files from the library $\mathcal{M}$. We denote by $q_{i,k}$ the probability that combination $i$ of requesting $k$ data items is selected. Suppose that combination $i$ has the indices $a_{i,k}=\{l_{i,1},l_{i,2},\cdots,l_{i,k}\}$. Hence, $q_{i,k}$ will be given by 
\begin{equation}
    q_{i,k} = \prod_{l \in a_{i,k}} p_v^{l} . \prod_{r\notin a_{i,k}} \Bigl(1-p_v^r\Bigr).
\end{equation}
The RSU considers all the vehicles which are located within its coverage area and considers its corresponding contact time. It also considers the selected combination of requested data items. Each data item in the requested combination will be delivered within $\frac{C}{\alpha_s} + \tau$ seconds. 
\subsection{Proactive Network}
In proactive network, each RSU exploits its limited memory to proactively cache some of the data items. For a RSU to serve a vehicle request from its cache, it requires $\frac{C}{\alpha_s}$ seconds. If the requested data item is not available in the RSU local cache, the data item will be delivered from the network backhaul within ($\frac{C}{\alpha_s} + \tau$) seconds. Since vehicle $v$ will be connected to the serving RSU $s$ for $h_v^s$ seconds, the vehicle can receive at most $\hat{M}_v^s$ data items from the RSU local cache where
\begin{equation}\label{Eq:M_Hat}
    \hat{M}_v^s := \min \left( \left\lfloor{ \frac{h_v^s}{\frac{C}{\alpha_s}} }\right\rfloor, \sum_{m=1}^M x_s^m \right).
\end{equation}
Since $\hat{M}_v^s$ is the minimum between most amount of cached files that can be received by vehicle $v$ and total cached items. If the contact time allows the vehicle to receive more data, the RSU will be able to deliver at most $\tilde{M}_v^s$ data items from the network backhaul where
    \begin{equation}\label{Eq:M_Tilde}
        \tilde{M}_v^s := \left\lfloor{ \frac{h_v^s - \hat{M}_v^s \frac{C}{\alpha_s}   }{\frac{C}{\alpha_s} + \tau} }\right\rfloor.
    \end{equation}
    Since the maximum of $\hat{M}_v^s$ is $\left\lfloor{ \frac{h_v^s}{\frac{C}{\alpha_s}} }\right\rfloor$ which means that content items can by delivered  from the backhaul  $ \tilde{M}_v^s$ equals zero.
Now, each vehicle can receive at most $\Bar{M}_v^s$ data items but can not exceed the library size $M$, where
    \begin{equation}
        \Bar{M}_v^s :=\min \left( \hat{M}_v^s + \tilde{M}_v^s , M \right) .
    \end{equation}
Therefore, the expected total delay incurred by all vehicles for RSU $s$ will be given by 
    \begin{equation}\label{Eq:Proactive_Time}
        \begin{aligned}
            W_{s}^\mathcal{P} &= \sum_{v=1}^{V} \left( \theta_v^{s} \sum_{k=0}^{\Bar{M}_v^s} \sum_{i=1}^{G_k} q_{i,k} t_{i,k}\right),
        \end{aligned}
    \end{equation}
where $\mathcal{P}$ represents the proposed proactive network, and $t_{i,k}$ is the transmission time for combination $i$. In general,
\begin{equation}
 {t_{i,k}}=
 k \left(\frac{C}{\alpha_s}+\tau \right)-\tau \sum_{l \in a_{i,k}} {x_s^l} , 
\end{equation}
where the first term represents total time if the user is served directly from the backhaul and the second term refers to saved time due to the cached files. We aim to minimize the latency over all users under the proactive network by finding an optimal caching policy $x_s^{m*},\forall m \in \mathcal{M}, s \in \mathcal{S}$. To evaluate the potential gains of proactive caching and its effect on the expected total delay, we characterize caching gain $\Delta W_s :=    \frac{W_s^{\mathcal{R}}}{\sum_{v=1}^{V} M_v^s} - \frac{W_s^{\mathcal{P}*}}{\sum_{v=1}^{V}\Bar{M}_v^s}$ as our performance metric for the proposed model as the difference between the expected time delay per file at reactive and proactive models.



\section{Non-cooperative Caching Scheme}\label{Sec:NonCooperative}

In this scheme, each RSU finds its optimal caching decision regardless of the other RSUs decisions. We assume that there is no cooperation nor coordination between the RSUs. Therefore, the optimization problem for RSU $s$ can be defined as follows
    \begin{equation}\label{Eq:OptProblem_NonCop}
        \begin{aligned}
            &  \underset{x_{s}^m}{\min}   \hspace{5mm} \frac{W_{s}^\mathcal{P}}{\sum_{v=1}^{V}\Bar{M}_v^s}  + \gamma \sum_{m=1}^{M} x_{s}^m \\
            & \text{s.t.} \hspace{6mm} (\ref{Eq:Caching_Const_1}), (\ref{Eq:Caching_Const_2}),
        \end{aligned}
    \end{equation}
where $\gamma$ is a factor capturing the caching cost per data item at any RSU. The first term represents the expected time per file and the second term refers to the total caching cost. We aim to minimize the expected time and cost while maximizing the total number of files that users can receive.
Due to the complexity of the problem, it is not easy to find the global optimal solution $x_{s}^{m*}$ of (\ref{Eq:OptProblem_NonCop}) as it is difficult to prove the convexity of the objective function. To overcome this difficulty, we find the optimal solution by launching an exhaustive search algorithm to attain the optimal combination of files that the RSU should cache. Therefore, we calculate the total expected time (main operation at the algorithm) by going over all different possible combinations of cached items from the library $\mathcal{M}$ to give $2^M$ iterations. Moreover, we take into consideration different demand combinations over all users which lead to $\sum_{v=1}^{V} \sum_{i=1}^{\Bar{M}_v^s} \binom{M}{i}$ operations. Hence, the complexity of the brute force algorithm is $\mathcal{O} \left(2^M \sum_{v=1}^{V} \sum_{i=1}^{\Bar{M}_v^s} \binom{M}{i}\right)$. It is clear that the complexity of the optimization problem increases exponentially as the number of vehicles and data items increases. Therefore, we propose an efficient greedy algorithm to tackle this issue. The key of this greedy algorithm is the term $\sum_{v=1}^{V} \theta_v^{s} p_v^m$ in (\ref{Eq:Proactive_Time}), which is considered as a weight factor for each data item. Moreover, the effect of vehicles velocity should be considered in the system by defining $\lambda_v^s:=\frac{h_v^{s}}{\sum_{v=1}^{V}h_v^{s}}$. This factor captures the vehicles velocity and gives high priority for low speed vehicles aiming to maximize the network throughput. 

The data item which receives high values of $\sum_{v=1}^{V} \theta_v^s p_v^m$ and $\lambda_v^s$ will be more probable to be cached first. Therefore, by sorting the data items based on these values, the RSU fills its local cache memory until it reaches the minimum between its cache size $Z_s$ and $\left\lfloor \alpha_s  \Bar{\mathbf{h}}^s \right\rfloor$, which refers to the average number of cached items received by users due to their mobility, since $\Bar{\mathbf{h}}^s$ is the average of  contact time vector. This allows us to guarantee that the caching cost stays reasonable and the caching process is efficient. The proposed algorithm gives promising results when compared to the exhaustive search as shown in Section \ref{Sec:Results}. It achieves lower complexity  $\mathcal{O}(MV+M^2)$, where $(MV)$ is the iterations for finding $\pi_s^m$ and $M^2$ is the complexity of the sorting algorithm \cite{paper_15}. Algorithm \ref{Alg:NonCoop_Suptimal} states the details of the proposed sub-optimal caching policy for the non-cooperative scheme.
\begin{algorithm}[h!]
\SetAlgoLined
    Initialize $p_v^m,\theta_v^s, h_v^s, \forall v \in \mathcal{V}, \forall m \in \mathcal{M}$\\
    Set $x_s^m = 0, \forall m \in \mathcal{M}$\\
    Evaluate $\pi_s^m := \sum_{v=1}^{V} \theta_v^s p_v^m \lambda_v^s, \forall m \in \mathcal{M}$ \\
    Sort $\pi_s^m$ descendingly and save sorting indices in $\zeta_s$\\
    Set $m = 1$\\
    \While{$C \sum_{m=1}^{M} x_{s}^m\leq \min\left(Z_s,\left\lfloor  \alpha_s  \Bar{\mathbf{h}}^s \right\rfloor  \right) $}
    {
        Set $x_s^{\zeta_s(m)} = 1$ \\
        Set $m = m+1$
    }
    Output $x_s^m$.
\caption{Sub-optimal noncooperative caching policy}
\label{Alg:NonCoop_Suptimal}
\end{algorithm}

\section{Co-operative Caching Scheme}\label{Sec:Cooperative}

In this scheme, we assume that the RSUs cooperate together through information signals to enhance the user experience by caching files which are expected to be requested at the coverage area of the next RSU. Therefore, we can rewrite (\ref{Eq:reactive_time}) as follows
 
      \begin{equation}\label{Eq:reactive_time_2}
        W_{s}^\mathcal{R} = \sum_{v=1}^{V} \left( \hat{\theta}_v^{s}\sum_{k=0}^{M_v^s} \sum_{i=1}^{G_k} \hat{q_{i,k}} k \left( \frac{C}{\alpha_s} + \tau \right)\right),
    \end{equation}
    
    \begin{equation}
           \hat{\theta}_v^{s}= \begin{cases} 1,&\text{$v$ was previously located at RSU $s-1$},\\
                             0, &\text{otherwise},
                               \end{cases}
    \end{equation}
where $\hat{\theta}_v^{s}$ is the updated location probability which becomes deterministic and we denote to previous RSU by $s-1$. $\hat{q}_{i,k} $ is the updated demand probability since the prior RSU passes the information about the files delivered to each user. Therefore, the current RSU $s$ updates $\vec{p}_{v}$ to $\hat{\vec{p}}_{v}$ by making $p_v^m=0, \hspace{3mm} \forall m \in \hat{ \mathcal{M}}_v^{s-1}$, where $\hat{ \mathcal{M}}_v^{s-1}$ is the set files downloaded by vehicle $v$ at the coverage area of RSU $s-1$, which means that the user will not request a file that had been received. By finding conditional distribution on previous demand of probability vector, we keep $ \sum_{m=1}^{M}\hat{p}_{v}^{m}=1$. Moreover, the total expected delay under the proactive caching scheme for RSU $s$ is given by
     \begin{equation}\label{Eq:Proactive_Time_2}
        \begin{aligned}
            W_{s}^\mathcal{P} &= \sum_{v=1}^{V} \left(\hat{\theta_v^{s}} \sum_{k=0}^{\Bar{M}_v^s} \sum_{i=1}^{G_k} \hat{q_{i,k}} t_{i,k}\right).
        \end{aligned}
    \end{equation}
For simplicity, we consider two RSUs only and then generalize the solution for all RSUs. In this case, the optimization problem for the two RSUs $s$ and $s-1$ is
	\begin{equation}\label{Eq:Optimization_Problem_2}
        \begin{aligned}
            &\underset{x_{s}^m,x_{s-1}^m}{\min}  \hspace{0mm} \frac{W_{s}^\mathcal{P}}{\sum_{v=1}^{V}\Bar{M}_v^s}+\frac{W_{s-1}^\mathcal{P}}{\sum_{v=1}^{V}\Bar{M}_v^{s-1}}\\
            &\hspace{10mm} + \gamma \sum_{m=1}^{M}\left( x_{s}^m+x_{s-1}^m\right)\\
            & \text{s.t.} \hspace{10mm} (\ref{Eq:Caching_Const_1}), (\ref{Eq:Caching_Const_2}).
        \end{aligned}
    \end{equation}
Inspired by the solution of the non-cooperative caching scheme discussed in Algorithm \ref{Alg:NonCoop_Suptimal}, we propose a sub-optimal caching policy for the cooperative caching scheme. RSU $s-1$ finds its optimal caching policy based on Algorithm \ref{Alg:NonCoop_Suptimal}. RSU $s$ modifies the calculation of $\pi_s^m$ to capture the updated information about the vehicles location and demand from RSU $s-1$ and proceeds with the same approach as in Algorithm \ref{Alg:NonCoop_Suptimal}. In particular, RSU $s$ fills its cache memory with the updated information about the location and demand of the passing vehicles after being served by RSU $s-1$. Algorithm \ref{Alg:Coop_sub-optimal} shows the details of the proposed sub-optimal caching policy under the cooperative scheme for two RSUs. The same approach can be applied to generalize the proposed solution for all RSUs. The caching decision of RSU $s$ will be affected by the decisions of RSUs $1,2,\cdots,s-1$. 
\begin{algorithm}[h!]
\SetAlgoLined
    Initialize $p_v^m,\hat{p}_v^m,\theta_v^{s-1}, \hat{\theta}_v^{s}, h_v^{s-1}, h_v^s \hspace{5mm}$ \\
    $\forall v \in \mathcal{V}, \forall m \in \mathcal{M},\forall s-1,s \in\mathcal{S}$\\
    Set $x_{s-1}^m = 0, \forall m \in \mathcal{M}$\\
    Evaluate $\pi_{s-1}^m := \sum_{v=1}^{V} \theta_v^{s-1} p_v^m \lambda_v^{s-1}, \forall m \in \mathcal{M}$ \\
    Sort $\pi_{s-1}^m$ descendingly and save sorting indices in $\zeta_{s-1}$\\
    Set $m = 1$\\
   \While{$C \sum_{m=1}^{M} x_{s-1}^m\leq \min\left(Z_{s-1},\left\lfloor  \alpha_{s-1} \Bar{\mathbf{h}}^{s-1} \right\rfloor  \right) $}{
        Set $x_{s-1}^{\zeta_{s-1}(m)} = 1$\\
        Set $m = m+1$
    }
    Set $ x_s^m = 0, \forall m \in \mathcal{M} $\\
    Evaluate $\pi_s^m := \sum_{v=1}^{V} \hat{\theta}_v^{s} \hat{p}_v^m \lambda_v^s, \forall m \in \mathcal{M} $ \\
    Sort $\pi_s^m$ descendingly and save sorting indices in $\zeta_s$\\
   \While{$C \sum_{m=1}^{M} x_{s}^m\leq \min\left(Z_s,\left\lfloor  \alpha_s  \Bar{\mathbf{h}}^s \right\rfloor  \right) $}{
        Set $x_s^{\zeta_s(m)} = 1$\\
        Set $m = m+1$
    }
    Output $x_{s-1}^m,x_s^m$.
 \caption{Sub-optimal Cooperative Caching Policy}
 \label{Alg:Coop_sub-optimal}
\end{algorithm}

\section{Numerical Results and Discussion}\label{Sec:Results}

In this section, we compare the performance of the proposed model to the reactive baseline model. 
We consider the case of two RSUs with equal coverage distances $L_s = 50$ meters and are serving 3 vehicles. We assume that the vehicles are interested in $M = 20$ data items each of size $C=1000$ kbytes. Vehicles velocities are generated by a truncated Gaussian distribution with mean $\mu=55$, variance $\sigma^{2}=10$, minimum velocity $u_{min} = 10$ km/h and maximum velocity $u_{max} = 120$ km/h. 
Both RSUs have the same data rate $\alpha_s = 1000$ kbyte/s, $\forall s \in \mathcal{S}$. For simplicity, we consider $\tau = C/\alpha_s = 1$ second and $\gamma$ has three different values $\{0.1,0.01,0.001\}$. 
We assume that each user has different interests from others. We model the demand probabilities for the users using Zipf distribution with different parameters. Moreover, to guarantee that the files demanded don't have the same order at all users, we permute the generated files ranks randomly to attain a different demand vector for each user.

\subsection{Non-cooperative versus cooperative caching schemes}

\begin{figure}[h!]
    \centering
\includegraphics[width=0.53\textwidth]{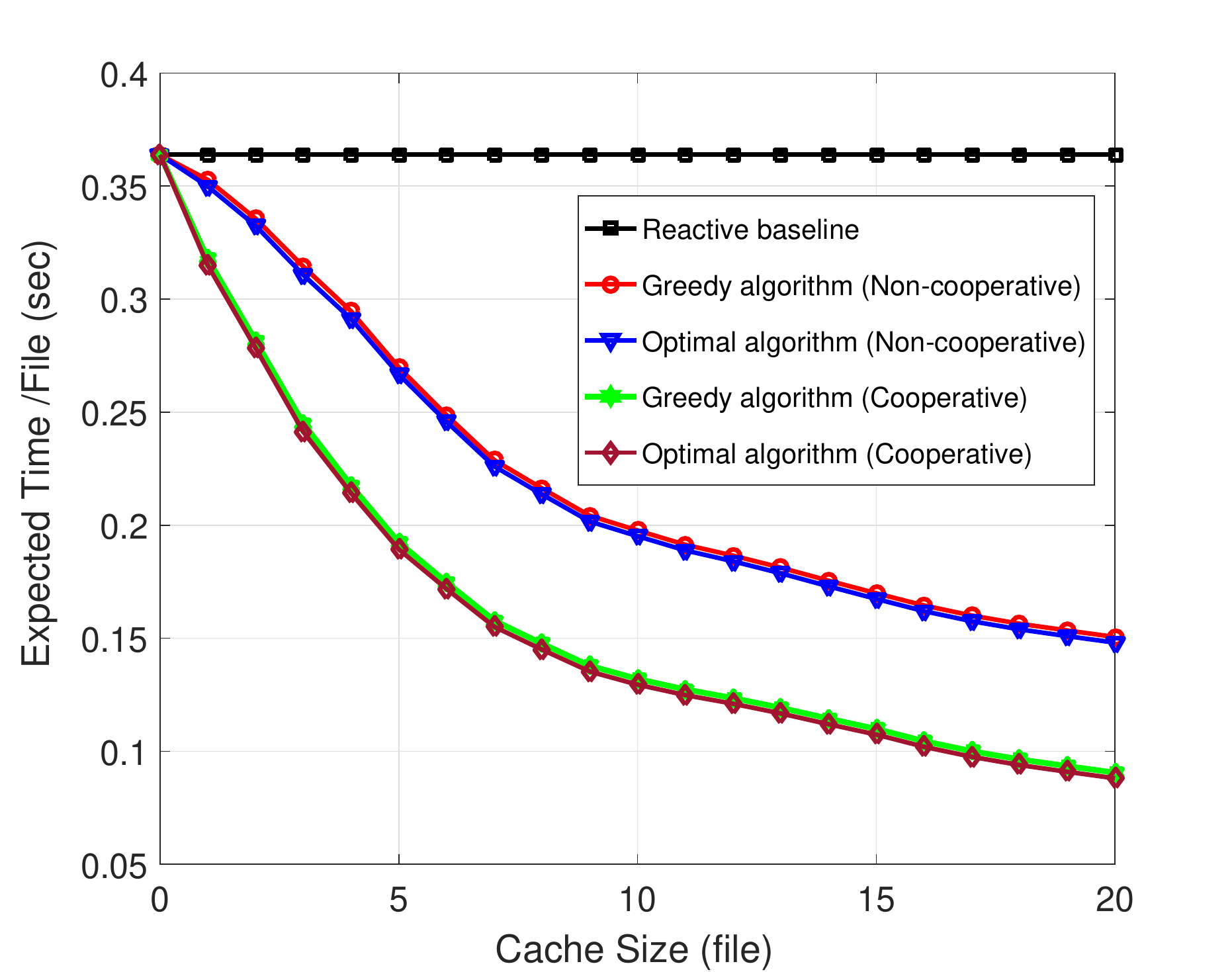}
    \caption{Comparison between the performance of reactive, optimal non-cooperative and optimal cooperative  caching schemes and proposed greedy algorithms .}
    \label{fig:CoopNonCoop}
\end{figure}

Fig. \ref{fig:CoopNonCoop} shows the expected time per file for different schemes versus the cache size $Z_s$. It is clear that the reactive baseline scenario has the worst performance due to absence of caching. In case of non-cooperative caching scheme, each RSU takes its decision independently, neglecting previous information about demand and location statistics. The figure shows that the expected time decreases as the RSU cache size increases to give a high caching gain $\Delta W_s$ of 45\% at cache size $=10$ files. On the other hand, in case of the cooperative caching scheme, each RSU collaborates with the previous RSUs to update its information about the vehicles' mobility and demand patterns. This collaboration allows the RSUs to increase their certainty about users behaviour and enhance their caching decisions. It is noticeable that the cooperative caching scheme outperforms the non-cooperative caching scheme for all cache sizes. For instance, the cooperative caching scheme achieves caching gain of $64\%$ at cache size $=10$ files. 
Moreover, the proposed sub-optimal greedy algorithm achieves nearly the same performance as the high complexity optimal exhaustive search algorithm.
\subsection{Impact of Caching cost on the system}
\begin{figure}[h!]
    \centering
\includegraphics[width=0.53\textwidth]{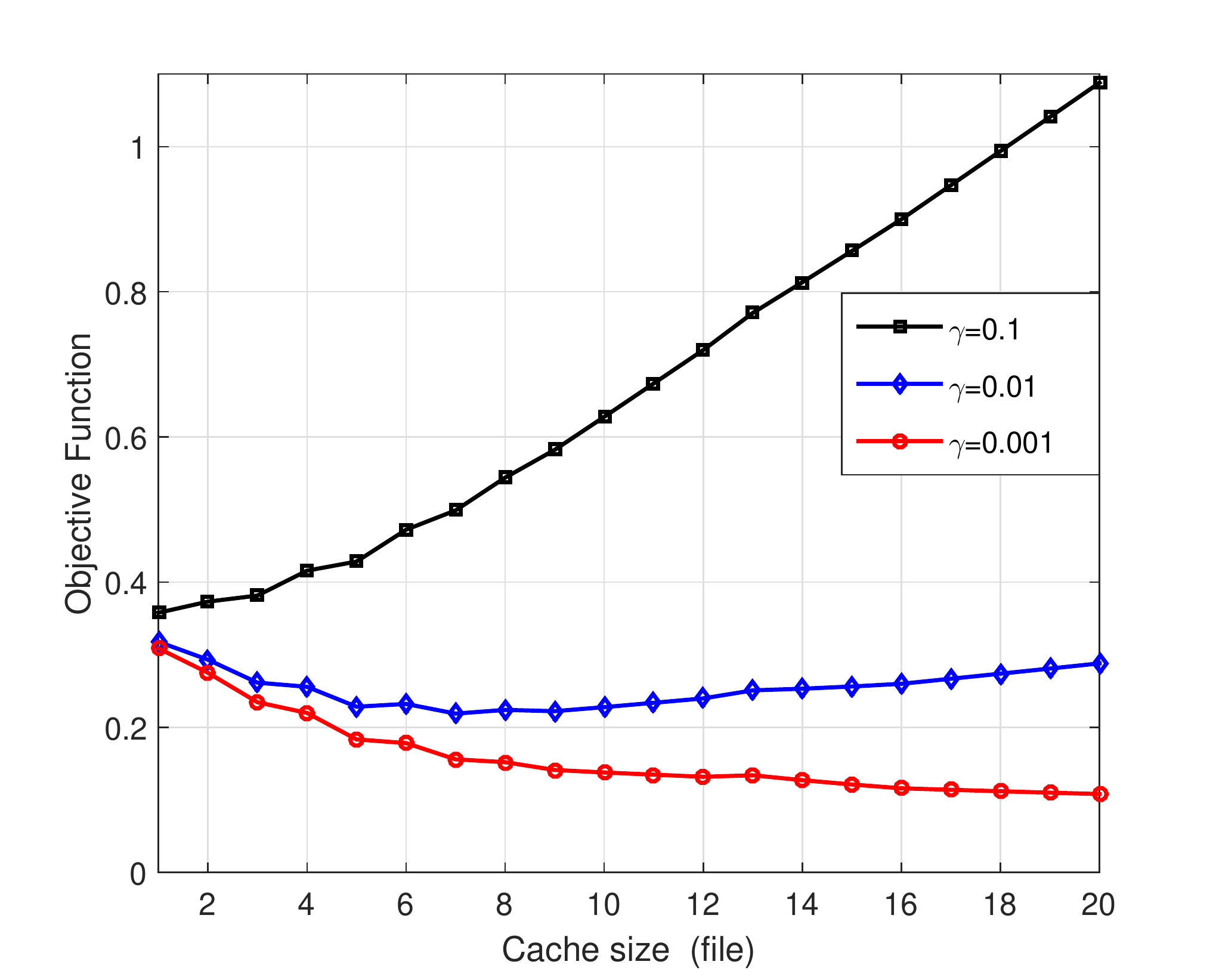}
    \caption{Impact of different cost factors in case of non-cooperative caching scheme.}
    \label{fig:cost}
\end{figure}

We characterize our model by considering that caching a file at the RSU causes some sort of cost such as time, energy or memory, etc. In our model, $\gamma$ is the factor that captures the caching cost. In Fig. \ref{fig:cost}, we plot the objective function at non-cooperative caching scheme (\ref{Eq:OptProblem_NonCop}) versus the cache size for different values of $\gamma$. At a high value of caching cost factor, $\gamma=0.1$, the minimum value of total cost is achieved at cache size $= 1$ file, which indicates that the optimal decision is to cache one file. While at intermediate values of caching cost factor, at $\gamma=0.01$, we get minimum total cost value at cache size $=7$ files, which is approximately the average number of cached files that users can receive$\left\lfloor  \alpha_s  \Bar{\mathbf{h}}^s \right\rfloor $, since RSU doesn't need to cache beyond this number to avoid any extra cost. At low caching cost factor, $\gamma=0.001$, the optimal decision is to cache all of the files in the library. Caching cost factor value may differ from time to time as it depends on abundant factors such as available memory, battery lifetime, and even peak and off-peak times.

\section{Conclusion}\label{Sec:Conclusion}

We study the gains achieved by proactive caching in RSUs to minimize the communication latency in VANETs. Information about the user demand and mobility is harnessed to cache a number of files at the RSUs before the actual demand. We capture the vehicles velocity in our model, which follows Gaussian distribution, to study its effect. We evaluate the proposed model by considering the expected network delay as a performance metric. Optimization problems are formulated to minimize the expected network delay for non-cooperative and cooperative caching schemes. To overcome the complexity of these problems, we propose an efficient greedy algorithm for each scheme. We compare the performance of the optimal and sub-optimal caching policies. Moreover, we investigate the effect of using different caching cost factors on its performance. Results reveal that the cooperative caching scheme outperforms the non-cooperative caching scheme for all cache sizes. Furthermore,  proactive caching is proven to be highly efficient at vehicular network when compared to the baseline reactive scheme. 

In this work, we considered a single highway road to study the effect of proactive caching on the network performance. We plan to extend our work to consider a complete city model consisting of multiple roads. We expect to achieve more gains by considering larger network size. Moreover, considering the mobility of vehicles across multiple roads may help us to address proactive caching more efficiently.

\bibliographystyle{IEEEtran}



\end{document}